# Comment: Citation Statistics

**Peter Gavin Hall**

I remember a US colleague commenting, in the mid 1980s, on the predilection of deans and other university managers for assessing academic statisticians' performance in terms of the numbers of papers they published. The managers, he said, "don't have many skills, but they can count." It's not clear whether the management science of assessing research performance in universities has advanced greatly in the intervening quarter century, but there are certainly more things to count than ever before, and there are increasingly sophisticated ways of doing the counting.

The paper by Adler, Ewing and Taylor is rightly critical of many of the practices, and arguments, that are based on counting citations. The authors are to be congratulated for producing a forthright and informative document, which is already being read by scientists in fields outside the mathematical sciences. For example, I mentioned the paper at a meeting of the executive of an Australian science body, and found that its very existence generated considerable interest. Even in fields where impact factors, $h$-factors and their brethren are more widely accepted than in mathematics or statistics, there is apprehension that the use of those numbers is getting out of hand, and that their implications are poorly understood.

The latter point should be of particular concern. We know, sometimes from bitter experience, of some of the statistical challenges of comparing journals or scientists on the basis of citation data—for example, the data can be very heavy-tailed, and there are vast differences in citation culture among different areas of science and technology. There are major differences even within probability and statistics. However, we have only rudimentary tools for quantifying this variation, and that means that we can provide only limited advice to people who are using citation data to assess the work of others, or who are themselves being assessed using those data.

Therefore, one of the conclusions we should draw from the study by Adler, Ewing and Taylor is that we need to know more. Perhaps, as statisticians, we could undertake a study, possibly funded in part by a grant awarding agency or our professional societies, into the nature of citation data, the information they contain, and the methods for analysing them if one must. This would possibly require the assistance of companies or organizations that gather such data, for example, Thomson Reuters and the American Mathematical Society. However, without a proper study of the data to determine its features and to develop guidelines for people who are inevitably going to use it, we are all in the dark. This includes the people who sell the data, those who use it to assess research performance and those of us whose performance is judged.

It should be mentioned, however, that too sharp a focus on citation analysis and performance rankings can lead almost inevitably to short- rather than long-term fostering of research excellence. For example, the appropriate time window for analyzing citation data in mathematics and statistics is often far longer than the two to three years found in most impact factor calculations; it can be more like 10–20 years. However, university managers typically object to that sort of window, not least because they wish to assess our performance over the last few years, not over the last decade or so. More generally, focusing sharply on citations to measure performance is not unlike ranking a movie in terms of its box-office receipts. There are many movies, and many research papers, that have a marked long-term impact through a complex process that is poorly represented by a simple average of naive criteria. More-over, by relying on a formulaic approach to measuring performance we act to discourage the creative young men and women whom we want to take up research careers in statistical science. If they enjoyed being narrowly sized and measured by bean-counters, they'd most likely have chosen a different profession.

*Peter Gavin Hall is Professor of Statistics, University of Melbourne, Victoria, Melbourne, VIC 3010, Australia e-mail: halpstat@ms.unimelb.edu.au.*

To illustrate some of the issues connected with citation analysis I should mention recent experiences in Australia with the use of citation data to assess research performance. In the second half of 2007 the academies, societies and associations representing Australian academics were asked by our federal government to rank national and international journals, as a prelude to a national review of research and to the development of new methods for distributing "overheads" to universities. The request was not uniformly well received by the academic community. For example, I didn't like it. However, to the government's credit it did endeavor to consult. Different fields drew up journal rankings in four tiers, using methods (e.g., deliberation by committee) that they deemed appropriate. But the conservative government that proposed this process lost office in November 2007, and a month later the Labor government that replaced it quietly but assiduously set about revising the rankings. They still used four tiers, consisting of the top 5%, next 15%, next 30% and lower 50% of the cohort of journals in a given field. (Selecting the cohort was, and is still, a controversial matter.) However, in many cases the revised rankings differed substantially from the earlier ones.

In probability and statistics, and applied mathematics, the revised rankings were worked out by the bureaucracy and by consultants whom the government employed, using five-year journal impact factors apparently computed from purchased data. The resulting ranking departed from accepted norms in a number of important respects, enough to shed significant doubt on the credibility of the whole exercise. Initially the procedures laid down by the Australian Research Council (ARC) for commenting on their revised ranking seriously restricted the ability of the probability and statistics community to respond as a body, for example through a committee. However, thanks to timely intervention by the IMS President in early July 2008, we were given an opportunity to make a submission directly to the ARC.

This enabled us to form a committee to recommend the correction of a number of serious problems. For example, the ARC's revised ranking based on impact factors had dictated that no journals in probability could be in the top tier; probabilists generally publish less, and are cited less, than statisticians. Even within statistics there were a number of what I regarded as significant errors. For example, some high impact factor journals, dedicated to specific fields of application, were placed into much higher tiers than renowned journals that focused more on the development of general statistical methodology. Still other important journals were omitted entirely. The committee set to work to remedy these problems.

As you can imagine, the redistribution of journals among tiers was not without significant debate. I received very strong email messages from, for example, a medical statistician who objected strenuously to *Statistics in Medicine* being in a lower tier than the *The Annals of Probability.* As he pointed out, the committee revising the ranking had "no objective criterion" for journal ranking other than impact factors, and in Thompson Reuters' most recent (i.e., 2007) list of those factors, *The Annals of Probability* had an impact factor of only 1.270, whereas *Statistics in Medicine* enjoyed 1.547. Then there were the upset probabilists, who objected to the large number of statistics journals in the top tier, relative to the small number of probability journals. One probabilist suggested a substantial reduction in the number of statistics journals being considered. Several argued that too much attention was being paid to impact factors. (I was unsuccessful in persuading my statistics colleagues to move far enough away from an impact-factor view of the world to put the *The Annals of Applied Probability* into the top tier, but colleagues on the applied mathematics committee generously adopted the journal and placed it in their first tier.)

As these experiences indicate, the lack of a clear understanding by the probability and statistics community of the strengths and weaknesses of citation analysis is causing more than a few problems. If the Australian government has its way, whether a paper is published in a first- or second-tier journal will influence the standing of the associated research, and will affect the "overhead" component of funding that flows to a university in connection with that work. I think this is quite wrong, but at present we do not have much choice other than to make the best of a bad deal. In that context, if our community does not have a clear and authoritative understanding of the nature, and hence the limitations, of impact factors (and more generally of citation data), then we cannot react in an authoritative way to arguments that we feel are invalid, but are nevertheless strongly held. Frankly, we need to know more about citation data and citation analysis, and that requires investment so that we can investigate the topic.